\documentstyle[prl,aps]{revtex}
\input psbox
\input psfig
\begin{document}
\draft

\twocolumn[\hsize\textwidth\columnwidth\hsize\csname@twocolumnfalse%
\endcsname

\preprint{SU-ITP 96/30, cond-mat/9607014}

\title{
Quantum Hall quarks or Short distance physics of 
quantized Hall fluids}

\author{Martin Greiter}
\address{Department of Physics, Stanford University, Stanford, CA 94305,
greiter@quantum.stanford.edu}

\date{SU-ITP 96/30, cond-mat/9607014, July 2, 1996}

\maketitle

\begin{abstract}
In order to obtain a local description of the 
short distance physics of fractionally quantized Hall states
for realistic (e.g.\ Coulomb) interactions, I 
propose to view the zeros of the ground state wave 
function,  as seen by an individual
test electron from far away, as particles.  I then
present evidence in support of this interpretation,
and argue that the electron effectively decomposes
into quark-like constituent particles of fractional charge.
\end{abstract}

]


In this letter, I will argue that electrons in
fractionally quantized Hall fluids effectively decompose into smaller,
quark-like particles, which then bind together to form electrons. 
This is not to say that electrons cease to be the 
fundamental degrees of freedom in these systems---a 
quantum mechanical description of all the electrons in the
liquid is as complete as any description can be---but 
rather that the hierarchy of effective field theories
is reversed.
While we usually assume that 
constituent particles are more fundamental 
than composite particles---quarks are thought as more fundamental than
hadrons in the standard model, or electrons as more fundamental
than Cooper pairs in superconductors---fractionally 
quantized Hall liquids provide us with
an example where the composite particles, the electrons,
are fundamental while the smaller constituent particles, 
which I call {\it quantum Hall quarks}, are fictitious
or effective degrees of freedom induced by the
surrounding electron condensate.  

I wish to address myself to readers 
without detailed knowledge of quantized Hall fluids,
and will begin with a review of the long distance physics.



Most of our understanding of the fractionally quantized Hall effect is
based on a highly original trial wave function for the
ground state proposed by Laughlin~\cite{la}:
\begin{equation}
\Psi_m [z] =\prod_{i<j}^N \,(z_i-z_j)^m\,
\prod_{i=1}^N e^{-{1\over 4}eB |z_i|^2}\,.
\label{eq:lau}
\end{equation}
This wave function describes a circular droplet of an incompressible
electron fluid in a strong perpendicular magnetic field $B$.
The fact that
all the electrons live in the lowest Landau level constrains 
the wave function to an analytic function in the complex
particle positions $z=x+i y$ times a Gaussian; the Jastrow
factor $\prod (z_i - z_j)$ raised to an odd integer power $m$
very effectively suppresses 
unwanted configurations in which electrons
come close to each other.

The Landau level filling fraction is defined as 
$\nu \equiv {\partial N}/{\partial N_\Phi}$,
where N is the number of electrons and $N_\Phi$ 
the number of Dirac flux quanta through the liquid. The latter
is equal to the number of zeros of the wave function 
$\Psi [z]$ seen by an individual test electron
with coordinate $z_1$ while all the other electron coordinates
$z_2,...,z_N$ are held at fixed positions.  
For the Laughlin state (\ref{eq:lau})
above, such a test electron will see $m$ zeros at the positions 
of each other electron, and no additional zeros elsewhere.
This implies $\nu =1/m$.

The elementary excitations, quasiholes and quasielectrons,
correspond to 
additional zeros which are not attached to electrons, 
or of deficits of zeros in given regions, respectively.
Laughlin's explicit trial wave function for the quasi-hole 
is given by
\begin{equation}
\Psi_m^\eta\, [z] = 
\prod_{i=1}^N \,(z_i-\eta )\ \Psi_m [z]\, .
\label{eq:lauqh}
\end{equation}
It is immediately obvious that $m$ quasiholes at the same point
$\eta $ amount to a true hole in the liquid, which has charge
$+e$; the convention here is $e>0$.  The quasihole charge is
therefore $e/m$.  There is a similar trial wave function for the
quasielectron, which involves derivatives in the $z_i$'s.
 
The trial wave function (\ref{eq:lau}) is actually
a rather good approximation to the exact ground state of two
dimensional electrons with Coulomb interactions 
in the lowest Landau level; at $\nu =1/3$,
a numerical comparison for 6 electrons on a sphere 
yields~\cite{fa} 
\begin{equation}
\langle\, \Psi_{m=3} \mid \Psi_{\rm exact}\,\rangle\,=\,0.9964\, .
\label{eq:lover}
\end{equation}
The reason for this remarkable agreement, or more 
generally for the success of Laughlin's theory, is that it
captures the correct long distance physics.  The essential
physics contained in the trial wave function (\ref{eq:lau})---in
fact the only physics except for the magnetic field---is 
that the electrons become {\it superfermions} for $m=3,5,...$etc.
The notion of superfermions 
makes sense in two space dimensions only.  It means
that the phase picked up by the wave function when one electron
encircles another is not $2\pi $, as Fermi statistics requires it,
but an odd multiple $2\pi m$,
which is consistent with Fermi statistics as well.  
The fractional quantum numbers of the quasiparticles, for example,
are a direct consequence of the superfermions.

Before closing this review, I would like to point out a technical 
detail~\cite{hald} 
which will ease the exposition below.
In the lowest Landau level, any two-body potential can be
parameterized by a discrete set of pseudopotentials $V_l$, which 
denote the energy cost of
having relative angular momentum $l$ between two particles.
The Laughlin 1/3 state is the exact ground state of 
a model Hamiltonian where only the pseudopotential $V_1>0$ while
all the other $V_l=0$ for $l=3,5,...$etc.
The reason for this is simply that the superfermions have---it follows
directly from their definition---no 
amplitude to be in a state of relative angular momentum $l=1$.





Now imagine we adiabatically deform this set of pseudopotentials
into the corresponding set for
Coulomb interactions.  Then the
ground state will evolve from a Laughlin $1/3$ state into
the exact Coulomb ground state at $\nu = 1/3$.  We know
from the overlap (\ref{eq:lover}) that the state cannot
change very much, and from the correctness of Laughlin's theory
that the long distance physics cannot change
at all---the changes must occur at short distances.  
The superfermions must evolve into {\it approximate superfermions},
that is, particles which look like superfermions from far away,
yet are different from the exact superfermions contained in Laughlin's
trial wave function.

\begin{figure}[hbt]
\psboxto(\columnwidth;0pt){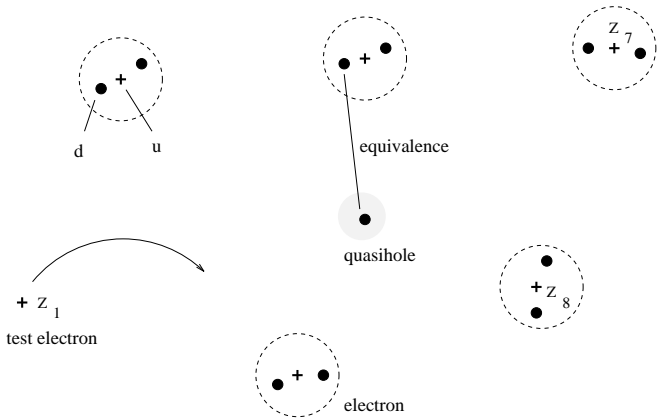}
\vspace{10pt}
\caption{Zeros of $\Psi_{\rm Coul.} [z]$ as seen by
an individual test electron $z_1$.  The zeros denoted by crosses
stem from a Jastrow factor and coincide with the electron positions
$z_2,...,z_N$, while those denoted by dots are in general very 
complicated functions of all the electron coordinates in the vicinity.
Also shown is an isolated zero not associated with any electron,
which corresponds to a quasihole excitation.}
\label{fig:zero}
\end{figure}

To elucidate this notion, consider once more 
the zeros of the wave function as seen by an individual test 
electron $z_1$ while all the other electron 
coordinates $z_2,...,z_N$ are fixed.  
The exact Coulomb ground state is of the general form
\begin{equation}
\Psi_{\rm Coul.} [z] =\prod_{i<j}^N \,(z_i-z_j)\,\, 
P(z_1,...,z_N) \,
\prod_{i=1}^N e^{-{1\over 4}eB |z_i|^2}.
\label{eq:cou}
\end{equation}
The Jastrow factor must be present since 
$\Psi_{\rm Coul.} [z]$
is antisymmetric; $P(z_1,...,z_N)$ is, in general, a
complicated symmetric polynomial.  A cartoon 
of the zeros of $\Psi_{\rm Coul.}$ 
in a given region, as seen by a test electron
from far away, is shown in Figure \ref{fig:zero}.
There are three zeros associated with each electron:  one of
them (denoted by a cross)
stems from the Jastrow factor in (\ref{eq:cou}) and coincides
with the electron coordinate $z_i$; the other two (denoted by dots)
stem from the polynomial $P(z_1,...,z_N)$ and are, 
in general, very complicated 
meromorphic functions of all the electron coordinates in a 
range which depends on the range of the interaction 
potential~\cite{halp}.  
In the limit of the minimally short ranged
potential mentioned above, the positions
of these two zeros depend only on the coordinate of the electron
they are associated with---in fact, they coincide with
this coordinate: $P(z_1,...,z_N)$ becomes a Jastrow factor squared,
and the general ground state (\ref{eq:cou}) the Laughlin 
1/3 state.

The reason the test electron must be
far away is that the positions of the zeros associated with each
electron depend on the position of the other electrons nearby.
If we were to 
pick an electron nearby as a test electron, the zeros seen in this
region would be those seen by another test electron from far away
if the test electron nearby would not exist.  The positions of the 
zeros would therefore depend on which of the electrons nearby
we were to pick as a test electron.  
If we, however, choose an electron far away as the test electron, 
the positions of the zeros in the region nearby 
will not depend on our choice 
and an interpretation of the zeros as particle coordinates, as I will
advocate below, is conceivable.

This brings me to the heart of the matter.  In order to provide
a {\it local} description of the short distance physics of 
fractionally quantized Hall fluids, I propose to view 
the zeros associated with the electrons as particles.  The electron
in a $\nu= 1/3$ state effectively decomposes into three smaller 
constituent particles,
\begin{equation}
e^- \to udd
\label{eq:udd}
\end{equation}
where the $u$ and $d$ particles, or quantum Hall quarks, are 
the zeros due to the Jastrow factor and the polynomial 
$P(z_1,...,z_N)$, respectively, as shown in Figure \ref{fig:zero}.  
The $d$ particles are equivalent to quasiholes,
in the sense that a quasihole is nothing but an isolated $d$.
The charge of the $d$ must therefore be equal to the charge 
of the quasihole, which we know to be $+1/3$.
Since the vacuum or ground state is neutral on a 
level on which the quasihole assumes this charge, the total charge
of the $udd$ composite must be zero, which implies that the charge 
of the $u$ is $-2/3$. 

The remainder of this letter is devoted to motivating 
and elucidating this idea.  To begin with, I will use the
hierarchy of quantized Hall fluids~\cite{hald,rbl:ss,bert}  
to establish an interpretation
of the quasiparticles in quantized Hall fluids as particles.



The quasiparticle excitations of quantized Hall liquids,
quasielectrons and quasiholes, were originally conceived as
vortices\,\cite{la}, and are adequately interpreted as such
when a plateau in the Hall resistivity results from their 
localization by disorder.  There are situations, however,
where an alternative interpretation as quantum mechanical particles
is not only possible, but inevitable.  The hierarchy of quantized
Hall states provides us with an example:  the quasiparticles
{\it themselves} condense into a Laughlin-Jastrow type fluid,
and it is necessary to assign a wave function
to them in order to describe this condensation. 
More precisely, we write an $[m,+p]$ state, that is a $p$ daughter
state of quasihole excitations of an $m$ parent state, as~\cite{bob,martin}
\begin{eqnarray}
\Psi_{[m,+p]}\, [z] &=& \int D[\xi ,\bar\xi ] \
\Phi_m\,[\bar\xi]\,\times \nonumber \\
&\times &\prod_{k<l}^{N_1}\, (\xi_k - \xi_l)^{1\over m} \,
\prod_{k=1}^{N_1} e^{-{1\over 4m}eB |\xi_k|^2}\,\times \nonumber \\
&\times  
&\prod_{k=1}^{N_1}\,\prod_{i=1}^N \,(z_i - \xi_k)\,\,\,\Psi_m [z]
\label{eq:hier}
\end{eqnarray}
with the quasiparticle wave function
\begin{equation}
\Phi_m\,[\bar\xi] =
\prod_{k<l}^{N_1}\, (\bar \xi_k - \bar \xi_l)^{p+{1\over m}} \,
\prod_{k=1}^{N_1} e^{-{1\over 4m}eB |\xi_k|^2} 
\nonumber
\label{eq:phi}
\end{equation}
and $N_1=N/p$.  The two factors in the second line of 
(\ref{eq:hier}) serve to normalize the quasiparticle Hilbert 
space. 
The fact that we have to integrate over the quasiparticle coordinates
to obtain a wave function for electrons is entirely consistent
with their nature as quantum mechanical particles, as quantum
mechanical degrees of freedom always have to be integrated out
with a wave function as a measure whenever we wish to
calculate a meassureable quantity (e.g.\ a transition probability).  

The explicit trial wave function (\ref{eq:hier}), and its cousin
for the $[m,-p]$ state in which quasielectrons rather than quasiholes
condense, are excellent approximations to the exact Coulomb ground 
states; at $\nu =2/5$, the overlap for 
6 electrons on a sphere is~\cite{martin}
\begin{equation}
\langle\, \Psi_{[3,-2]} \mid \Psi_{\rm exact}\,\rangle\,=\,0.9995\, ,
\label{eq:hover}
\end{equation}
a number which compares favorably even with the Laughlin 1/m states.

The particle nature of the quasiparticles leads us to the question of
their origin, to the question of where new particles of fractional
charge may come from.  The answer is the obvious one, and
this is precisely why it is so hard to swallow:  The charges of the 
quasiparticles are parts of electron charges, and {\it the
quasiparticles themselves are parts of electrons}.  
In order for quasiparticle excitations to exist, the vacuum or
ground state must contain them already in a confined phase---the
vacuum must be a phase in which pieces of electrons bind 
together to form electrons~\cite{fw}.





Particle physicists usually establish the existence of new particles
by observing them as resonances in scattering experiments.
This is not possible for quantum Hall quarks, once because
the kinetic energy of all the particles involved is quenched due
to Landau level quantization, and the concept of time
does consequently not exist, but even more profoundly so 
because we invoke 
quantum Hall quarks to describe the vacuum, which trivially excludes
the possibility of scattering experiments.

Fortunately, there is a way around these problems.  While we
do not have a concept of real time, we can perform a Monte Carlo
simulation and monitor scattering events as particle configurations
evolve in Monte Carlo time.  Let me briefly review the technique:
a Monte Carlo simulation is a numerical method 
to approximate an integral over many variables with a 
probability $\rho $ as a measure.
Instead of integrating over the variables directly, we interpret
them as dynamical variables, and let them evolve in Monte Carlo time.  
This concept of time is discrete; at each step we randomly
pick one of the variables, and define a new configuration by randomly
choosing a new value 
for this variable according to a certain distribution, which is usually
taken as a Gaussian centered at the present value. 
Finally, we randomly decide whether to
update the configuration or not according to probabilities proportional to 
the measure $\rho $ for the new and for the 
present configuration, respectively.  The desired integral
is obtained by averaging the
integrand (not including the measure) over a long span in Monte Carlo time;
the approximation becomes exact as this span tends to infinity.
\begin{figure}[hbt]
\psboxto(\columnwidth;0pt){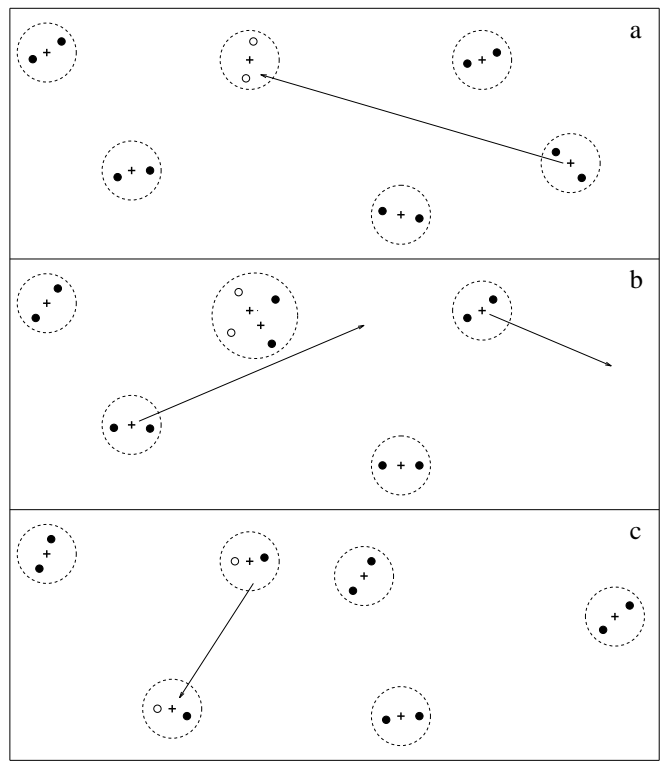}
\vspace{10pt}
\caption{Electron--electron scattering in Monte Carlo time:
a) two electron coordinates 
happen to come very close to each other.  
b) the surrounding electron configuration evolves in Monte Carlo time,
and with it the configuration of the zeros 
of the two electrons close to each other.  
c) the two electrons separate again, having interchanged one of 
their constituent particles.}
\label{fig:mont}
\end{figure}

In our case, 
the Monte Carlo variables are the electron coordinates 
$z_i$, and the measure $\rho $ is the probability 
$|\Psi_{\rm Coul.}(z_1,...,z_N)|^2$.  A snapshot of a typical 
Monte Carlo configuration
including all the zeros or quantum
Hall quarks, is shown in Figure \ref{fig:mont}a.  
Only the electron coordinates, or $u$ quantum Hall quarks, 
are truly dynamical variables
in Monte Carlo time; the dynamics of the remaining zeros, or
$d$ quantum Hall quarks, is induced through the surrounding 
electron condensate.  This, however, does not emerge from
Figure \ref{fig:mont}a, nor does it ever manifest itself as we
follow the evolution of this configuration on a continuous
time scale---that is, a time scale on which all the variables
evolve simultaneously.

Let us now look at a particular scattering event, as shown in 
Figure \ref{fig:mont}.  In this event, two electrons scatter 
off each other, and interchange one of their constituent particles:
two electron coordinates happen to come very close 
to each other, and remain unchanged for a number of Monte Carlo steps, 
while the configuration of the
additional zeros associated with them evolves
with the surrounding electron liquid; this configuration will,
in general, have changed significantly  
by the time the two electrons separate again.
Thus there is a finite amplitude for zeros to get interchanged---the 
zeros are {\it indistinguishable} when interpreted as particles, and
scatter into each other as identical particles do in quantum 
mechanics.  

This Monte Carlo experiment 
nicely illustrates the underlying
reason why it is possible for these fictitious or induced degrees of 
freedom to become particles:  induced
and fundamental degrees of freedom are {\it locally} equivalent, 
in the sense that 
no local experiment, and in particular no scattering experiment, is
capable of resolving the difference.  This is precisely the reason 
why it is perfectly reasonable to invoke quantum Hall quarks 
in order to provide a {\it local} description of fractionally 
quantized Hall fluids at short distances.



I have mentioned above that the
$d$ particle is equivalent to a quasihole excitation,
in the sense that a quasihole is nothing but a $d$ in isolation.  
To see this, we just need to perform another
Monte Carlo experiment with an exact quasihole for Coulomb
interactions at some location $\eta $, and we will find that
the position of the zero associated with the quasihole does not exactly 
coincide with the position~$\eta $, but rather depends on all the
electron coordinates in the vicinity, as indicated  
in Figure~\ref{fig:zero}.  Moreover, we will find that this zero
has a finite amplitude to get interchanged with 
other zeros or $d$ particle in the liquid as electrons 
scatter off the quasihole in Monte Carlo time.  This illustrates  
the precise sense in which the exact quasihole
for realistic interaction potentials differs from Laughlin's 
trial wave function (\ref{eq:lauqh}).
The equivalence of confined and isolated zeros can of course also
be deduced from the fact that a quasihole-quasielectron pair is 
created by removing a zero from the vacuum in a certain region and placing
it into another region.



Most of what I have explained in this letter concerns the ground
state or vacuum of fractionally quantized Hall fluids,  
while only excitations matter 
to experiments performed on quantum Hall systems.  The real
significance of the analysis presented here
lies in the general
message we can learn from it, and the potential relevance of this
message to other systems, in particular to the vacuum
of our universe, the ground state which supports all the elementary
particles known to us as excitations.

This general message is that some of the particles we see or detect as 
excitations above a certain vacuum 
might conceivably be pieces of larger particles invisible
to us.  The degrees of freedom we perceive as fundamental may in
fact be fictitious or induced, and fractional quantum numbers---but
in particular the fractional charges of quarks 
in quantum chromodynamics---may
arise through a mechanism related to the one responsible for
quantum Hall quarks.  If we specifically imagine an observer
who lives in a quantized Hall fluid and consists of quasiparticles,
this observer would never see electrons, but only fictitious particles of 
fractional charge, and would naturally be inclined to accept those
as fundamental.  Note in particular that scattering experiments,
both the ones performed by this observer as well as the ones
performed by us in particle accelerators, are incapable of 
resolving the ambiguity between induced and fundamental degrees of
freedom.

I wish to thank L.\ Alvarez-Gaum\'e, E.\ Br\'ezin, 
W.\ Krauth, R.B.\ Laughlin, L.\ Susskind, and E.\ Verlinde 
for inspiring discussions,
and to R.B.\ Laughlin again for his critical reading of the manuscript.
This work was supported through a Fellowship in 
Elementary Particle Physics at CERN and through NSF grant No.~DMR-95-21888.

\end{document}